\documentclass[12pt,preprint]{aastex}

\shorttitle{Magnetic Fields at the Base of the CZ}
\shortauthors{Chou \& Serebryanskiy}

\begin{document}

\title{Solar Cycle Variations of p-Mode Frequencies}

\author{Dean-Yi Chou and Alexander Serebryanskiy}

\affil{Institute of Astronomy and Department of Physics,
Tsing Hua University, \\
Hsinchu, 30043, Taiwan, R.O.C. \\
chou@phys.nthu.edu.tw, alex@astrin.uzsci.net}

\begin{abstract}
Observations show that the solar p-mode frequencies change with 
the solar cycle. 
The horizontal-phase-velocity dependence of the relative frequency change, 
scaled by mode mass,  
provides depth information on the perturbation in the solar interior.
We find that the smoothed scaled relative frequency change
varies along the solar cycle for horizontal phase velocities higher than
a critical value, which corresponds to a depth 
near the base of the convection zone.  
This phenomenon suggests that the physical conditions in a region near the 
base of the convection zone change with the solar cycle.

\end{abstract}


\section{Introduction}
Observations show that the solar p-mode frequencies vary with the solar cycle,
and the change in frequency correlates with magnetic activities. 
The previous study suggests that the measured frequency change is caused by the 
change in physical conditions near the surface \cite{lib90,shi91}, 
and no evidence of 
solar-cycle variations of physical conditions deep in the solar interior 
has been found from the measured frequencies 
\cite{how99, bas00, ant01, bas02, eff02, bas03}.
It is expected that the perturbation in the solar interior have a rather small 
contribution to the frequency change, if it exists, because of small 
$1/\beta$.  
It is also reasonable to expect, based on the current dynamo theories, that 
the second largest contribution to the frequency change is from 
the perturbation near the base of the convection zone (BCZ).
A perturbation near the BCZ would modify 
the frequencies of different modes in different ways: 
the frequencies of the modes which can penetrate into the BCZ would change, 
while the modes which can not penetrate into the BCZ would remain intact.
The penetration depth (lower turning point) of a mode is determined by 
the horizontal angular 
phase velocity $w$ $(\equiv \omega /L = \omega/[l(l+1)]^{1/2})$, 
where $\omega$ is the angular frequency and $l$ the mode degree.
Thus the contribution of the BCZ perturbation to the frequency change 
depends on $w$.
Here we present the evidence that the $w$ dependence of smoothed scaled 
relative frequency change, $(\delta \omega_{nl}/\omega_{nl}) E_{nl}$, 
varies with the solar cycle at large $w$, while remains constant at small $w$,
where $\delta\omega_{nl}$ is the frequency change and $E_{nl}$ the mode mass.

To search for the contribution of the near-BCZ perturbation to the frequency
change, we have to remove the contributions of the near-surface perturbation,
which is much greater than that of the BCZ perturbation.
In this study we adopt a method similar to one developed by 
Di Mauro et al. (2002) to remove
the near-surface contributions to solar-cycle variations of frequencies.
This method was originally developed to remove the contributions 
of uncertainties near the surface in solar models to the p-mode frequencies.
Here we use it to remove the contributions of the near-surface perturbation
caused by solar activities.

\section{Method}

In the asymptotic theory, the frequency change $\delta\omega_{nl}$,
due to sound-speed perturbations can be expressed as \cite{mauro02}
\begin{equation}
\frac{\delta \omega_{nl}}{\omega_{nl}} S_{nl}
= H_1(w) + H_2(\omega_{nl}) + H_3(\omega_{nl}, w) + H_4(\omega_{nl}, w)
\label{eq:h1234-a}
\end{equation}
where $S_{nl}$ is approximately proportional to the mode mass $E_{nl}$.
In this study we replace $S_{nl}$ by $\bar{E}_{nl}\equiv E_{nl}/E_{21,0}$, where
$E_{21,0}$ is the mode mass for mode $n=21$ and $l=0$.
In equation (\ref{eq:h1234-a}), 
$H_1$ is from the perturbations in the whole Sun, while
$H_2$, $H_3$, and $H_4$, are caused only by the near-surface perturbation.
Among the near-surface terms, $H_2(\omega_{nl})$, is the lowest order term.
It depends on frequency only, and could be expressed as \cite{mauro02}
\begin{equation}
H_2(\omega_{nl}) =  \sum_{i=1}^{M} a_i^{(2)} P_i(\tilde{\omega}_{nl}) 
\label{eq:h2-a}
\end{equation}
where $P_i$ are the Legendre polynomials, and $\tilde{\omega}_{nl}$ is  
the frequency scaled to be in the interval $[-1,1]$ for the modes
considered. 
The third and fourth terms, $H_3$ and $H_4$, are one order higher than
$H_2$ \cite{bro93,gou95}.  They are functions of $\omega_{nl}$ and $w$, 
and can be expressed as \cite{mauro02}
\begin{eqnarray}
H_3(\omega_{nl},w) &=& \frac{1}{w^2} \sum_{i=1}^{M} a_i^{(3)} 
P_i(\tilde{\omega}_{nl}) \nonumber \\ 
H_4(\omega_{nl},w) &=& \frac{1}{w^4} \sum_{i=1}^{M} a_i^{(4)} 
P_i(\tilde{\omega}_{nl}) 
\label{eq:h34-a}
\end{eqnarray}

The first term in equation (\ref{eq:h1234-a}), $H_1$, 
is from the perturbations in the whole Sun. 
It is a function of $w$ only, and can be expressed as an integral
\cite{chr88}
\begin{equation}
H_1(w) = \int_{r_t}^{R} \left( 1-\frac{c^2}{r^2 w^2} \right) ^{-1/2} 
\frac{\delta c}{c^2} \; dr
\label{eq:h1-a}
\end{equation}
where $r_t$ is determined by $r_t = c(r_t)/w$. 
The largest contribution to $H_1$ is from the near-surface perturbation.
The second largest contribution is expected from the perturbation near the BCZ. 
Here we define a critical horizontal phase velocity $w_b$ such that the modes
with $w<w_b$ are not affected by the BCZ perturbation, while the modes
with $w>w_b$ are affected by the BCZ perturbation in the ray theory.
It is reasonable to assume that the only non-negligible contribution to $H_1$
for the modes with $w<w_b$ is from the near-surface perturbation. 
If the near-surface perturbation is located within a thin layer at $r_0$, 
from equation (\ref{eq:h1-a}) $H_1$ can be approximated by  
\begin{equation}
H_1(w) \approx   s_0 \left( 1-\frac{w_0^2}{w^2} \right) ^{-1/2} 
\;\;\;\;\;\;\;\;\;\; {\rm for} \; w_0< w<w_b
\label{eq:h1-b}
\end{equation}
where $w_0 = c(r_0)/r_0$, and $s_0 = \int_{r_{bcz}}^R (\delta c/c^2) \, dr$.
The perturbation near the BCZ has a contribution to 
$(\delta \omega_{nl}/\omega_{nl}) \bar{E}_{nl}$ only for the range of 
$w>w_b$ through $H_1$. 
In the following analysis, we limit our study in a range of
$2.5 <\omega/2\pi < 3.5$ mHz  and $190 < w < 1570$, where 
$w$ is computed with frequency in units of $\mu$Hz.  
For this range, $(w_0/w)^2<10^{-2}$, if $r_0>0.993 R_{\odot}$.
Thus $H_1(w)$ from the near-surface perturbation can be further 
approximated by  
\begin{equation}
H_1(w) \approx   s_0 \left( 1+\frac{1}{2}\frac{w_0^2}{w^2} \right)
\equiv a_1^{(1)} + a_2^{(1)} \frac{1}{w^2} 
\;\;\;\;\;\;\;\;\;\; {\rm for} \; w<w_b
\label{eq:h1-c}
\end{equation}

The contributions from the near-surface perturbation can be expressed as 
\begin{equation}
\left[ \frac{\delta \omega_{nl}}{\omega_{nl}} \bar{E}_{nl} \right]_{\rm{surf}}
= a_1^{(1)} + a_2^{(1)} \frac{1}{w^2}
+ \sum_{i=1}^{M} a_i^{(2)} P_i(\tilde{\omega}_{nl}) 
+  \frac{1}{w^2} \sum_{i=1}^{M} a_i^{(3)} P_i(\tilde{\omega}_{nl}) 
+ \frac{1}{w^4} \sum_{i=1}^{M} a_i^{(4)} P_i(\tilde{\omega}_{nl}) 
\label{eq:surf-a}
\end{equation}
If we define $[(\delta \omega_{nl}/\omega_{nl}) \bar{E}_{nl}]_{\rm{bcz}}$ 
to be the contribution of the perturbations near and below the BCZ,
$(\delta \omega_{nl}/\omega_{nl}) \bar{E}_{nl}$ can be expressed as 
\begin{equation}
\frac{\delta \omega_{nl}}{\omega_{nl}} \bar{E}_{nl}=
\left[ \frac{\delta \omega_{nl}}{\omega_{nl}} \bar{E}_{nl} \right]_{\rm{surf}}
+ \left[ \frac{\delta \omega_{nl}}{\omega_{nl}} \bar{E}_{nl} \right]_{\rm{bcz}}
\label{eq:sum-a}
\end{equation}
where $[(\delta \omega_{nl}/\omega_{nl}) \bar{E}_{nl}]_{\rm{bcz}} = 0$ 
for $w<w_b$. 

To obtain $[(\delta \omega_{nl}/\omega_{nl}) \bar{E}_{nl}]_{bcz}$,
first, we fit equations (\ref{eq:surf-a}) to the measured 
$(\delta \omega_{nl}/\omega_{nl}) \bar{E}_{nl}$ for the range of $w<w_b$
to determine the coefficients, $a_i^{(1)}$,  $a_i^{(2)}$,  $a_i^{(3)}$,
and $a_i^{(4)}$.
Second, we use these coefficients to compute 
$[(\delta \omega_{nl}/\omega_{nl}) \bar{E}_{nl}]_{\rm{surf}}$ 
for all ranges of $w$.
Third, the contribution from the BCZ,
$[(\delta \omega_{nl}/\omega_{nl}) \bar{E}_{nl}]_{\rm{bcz}}$, is computed with
equation (\ref{eq:sum-a}).

\section{Modeling}

In this section, we test the above method with a model computation.
First, we perturb a standard model \cite{chr96}
by changing its adiabatic exponent
$\Gamma_1 (\equiv \partial \ln p/ \partial \ln \rho |_s )$ in two regions: 
one is located near the surface and another near the BCZ.
Each perturbation is a Gaussian in depth.
The near-surface perturbed region has a magnitude of 
$\delta\Gamma_1/\Gamma_1 = 5.95\times 10^{-3}$ located 
at $0.9995 R_{\odot}$ with a width of $10^{-3} R_{\odot}$. 
The magnitude of $\delta\Gamma_1/\Gamma_1$ is chosen such that
the frequency change is close to the measured value at maximum. 
The near-BCZ perturbed region has a magnitude of 
$\delta\Gamma_1/\Gamma_1 = 4\times 10^{-5}$ located 
at $0.713 R_{\odot}$ with a width of $0.05 R_{\odot}$.
Second, the mode frequencies of the perturbed model are computed
\cite{chr03}.
Third, the relative frequency difference between the perturbed 
and unperturbed models scaled by mode mass, 
$(\delta \omega_{nl}/\omega_{nl}) \bar{E}_{nl}$, is computed.
Its value versus $log(w)$ is shown in the upper left panel of Figure \ref{fig1}.
The scattering of the data is due to the frequency dependence.
It is noted that an increase in 
$(\delta \omega_{nl}/\omega_{nl}) \bar{E}_{nl}$ is visible 
around $log(w)=2.7$ which corresponds to the location
of the assumed perturbation near the BCZ.
To see the trend of the scattered data, we smooth data 
with a 41-point box running mean.
The advantage of smoothing is discussed in \S5.
The oscillatory features in the smoothed data are due to the
ridge structure of p-mode frequencies.  
Fourth, equation (\ref{eq:surf-a}) is fitted to  
$(\delta \omega_{nl}/\omega_{nl}) \bar{E}_{nl}$ for the range of
$w<w_b$ to determine the coefficients 
$a_i^{(1)}$,  $a_i^{(2)}$,  $a_i^{(3)}$, and $a_i^{(4)}$.
Here we adopt $w_b=430$ ($log(w)=2.63$), and $M=8$.
Fifth, these coefficients are used to compute 
$[(\delta \omega_{nl}/\omega_{nl}) \bar{E}_{nl}]_{\rm{surf}}$
with equation (\ref{eq:surf-a}) for all ranges of $w$.
Sixth, $[(\delta \omega_{nl}/\omega_{nl}) \bar{E}_{nl}]_{\rm{bcz}}$
is computed with equation (\ref{eq:sum-a}). 
The result is shown in the upper right panel of Figure \ref{fig1}.

The scattering of data is greatly reduced, especially for $w<w_b$.
It is because the frequency dependence, described by  
$[(\delta \omega_{nl}/\omega_{nl}) \bar{E}_{nl}]_{\rm{surf}}$, 
has been removed.  
The global trend of increase in 
$(\delta \omega_{nl}/\omega_{nl}) \bar{E}_{nl}$ with $w$ is also removed. 
Its $w$ dependence is flat and its value is close to zero.
For comparison, we plot 
$(\delta \omega_{nl}/\omega_{nl}) \bar{E}_{nl}$ 
computed with only the BCZ perturbation in the lower part of 
the upper right panel in Figure \ref{fig1}.
It can be seen that the near-surface contributions are reasonably well removed
and a small change in frequency caused by the near-BCZ perturbation 
is reasonably well recovered.

\section{Data Analysis and Results}

The solar p-mode frequencies used in this study are derived from the data 
taken with the Michelson Doppler Imager (MDI) on board the SOHO \cite{sch95}. 
The frequencies are measured with a time series of 72 days.
We use the modes in the range:  
$2.5 <\omega/2\pi < 3.5$ mHz  and $190 < w < 1570$,
because of no missing frequency in this range.
To avoid that the result may be dominated by a few frequencies  
with large errors, we discard those frequencies by the following
procedure.  First, for each mode, temporal variations of frequency is
fitted to a linear function.  Second, the standard deviation is computed.
Third, the frequencies deviated from the fit greater than three standard
deviation are discarded.  Less than $1\%$ of frequencies considered are
discarded. 

The frequency averaged over a solar minimum period, 1996.05 - 1997.07,
is used as the reference frequency.  The frequency change along the solar cycle
is computed relative to the reference frequency. 
The relative frequency change, scaled by mode mass, 
averaged over a maximum period (2000) is shown in the lower left panels of
Figure \ref{fig1}.  The data are rather scattered.
Part of scattering is due to the frequency dependence of 
$(\delta \omega_{nl}/\omega_{nl}) \bar{E}_{nl}$ as shown in the model
computation in \S2, part is due to the error of measured frequency.
The smoothed data with a 41-point box running mean as \S3 are shown with
a solid line.
The interesting phenomenon shown in the smoothed data is 
that the values of $(\delta \omega_{nl}/\omega_{nl}) \bar{E}_{nl}$ 
at small $w$ and large $w$ are different.
It is approximately constant for $log(w)<2.7$ and drops to a smaller 
value at $log(w) \approx 2.7$.

We use the method described in \S2 and \S3 to remove the contributions
of the near-surface perturbation to obtain  
$[(\delta \omega_{nl}/\omega_{nl}) \bar{E}_{nl}]_{\rm{bcz}}$.
The results are shown in the lower right panels of Figure \ref{fig1}.
Although the scattering of data is reduced, it is still
significantly greater than that in modeling.  It is probably caused by the 
the error of measured frequency.
The smoothed $[(\delta \omega_{nl}/\omega_{nl}) \bar{E}_{nl}]_{\rm{bcz}}$
is also shown by a solid line.
In the range of $w<w_b$, similar to the modeling in \S3, 
the global trend in $(\delta \omega_{nl}/\omega_{nl}) \bar{E}_{nl}$ is removed.
The $w$ dependence is rather flat and the value is about zero. 
The difference between small $w$ and large $w$ becomes more apparent in
$[(\delta \omega_{nl}/\omega_{nl}) \bar{E}_{nl}]_{\rm{bcz}}$.

To see how this phenomenon varies with the solar cycle, we plot the
smoothed $(\delta \omega_{nl}/\omega_{nl}) \bar{E}_{nl}$ 
for different periods along the solar cycle in the upper left panels of 
Figure \ref{fig2}.
In the period of low activity (1997-1998), the curve is rather flat.
As the activity increases, 
$(\delta \omega_{nl}/\omega_{nl}) \bar{E}_{nl}$ at $log(w) > 2.7$
becomes smaller relative to that at $log(w) < 2.7$.
The difference increases with the activity.
After subtracting the near-surface contributions, 
$[(\delta \omega_{nl}/\omega_{nl}) \bar{E}_{nl}]_{\rm{bcz}}$ is shown in
the upper right panel of Figure \ref{fig2}.
At $log(w)<2.7$, $[(\delta \omega_{nl}/\omega_{nl}) \bar{E}_{nl}]_{\rm{bcz}}$ 
is approximately flat and close to zero.  
At $log(w)>2.7$, $[(\delta \omega_{nl}/\omega_{nl}) \bar{E}_{nl}]_{\rm{bcz}}$ 
is negative.

\section{Discussion}

The $w$ dependence of $(\delta \omega_{nl}/\omega_{nl}) \bar{E}_{nl}$ 
provides depth information on the perturbation in the solar interior.
However, it is difficult to see the weak $w$ dependence of
$(\delta \omega_{nl}/\omega_{nl}) \bar{E}_{nl}$
from the scattered $(\delta \omega_{nl}/\omega_{nl}) \bar{E}_{nl}$. 
We smooth $(\delta \omega_{nl}/\omega_{nl}) \bar{E}_{nl}$ to  
search for its weak $w$ dependence. 
Smoothing has another advantage.
Smoothing is equivalent to averaging over frequencies at a fixed $w$.  
If the higher order terms, $H_3$ and $H_4$, are neglected 
in equation (1), smoothed $(\delta \omega_{nl}/\omega_{nl}) \bar{E}_{nl}$ 
equals to $H_1(w) + constant$, where the constant is the average of $H_2$
over frequencies which is independent of $w$. 
Thus smoothed $(\delta \omega_{nl}/\omega_{nl}) \bar{E}_{nl}$ 
gives the $w$ dependence of $H_1$.
This can be demonstrated with model computations. 

\begin{figure}[t]
\epsscale{0.65}
\plotone{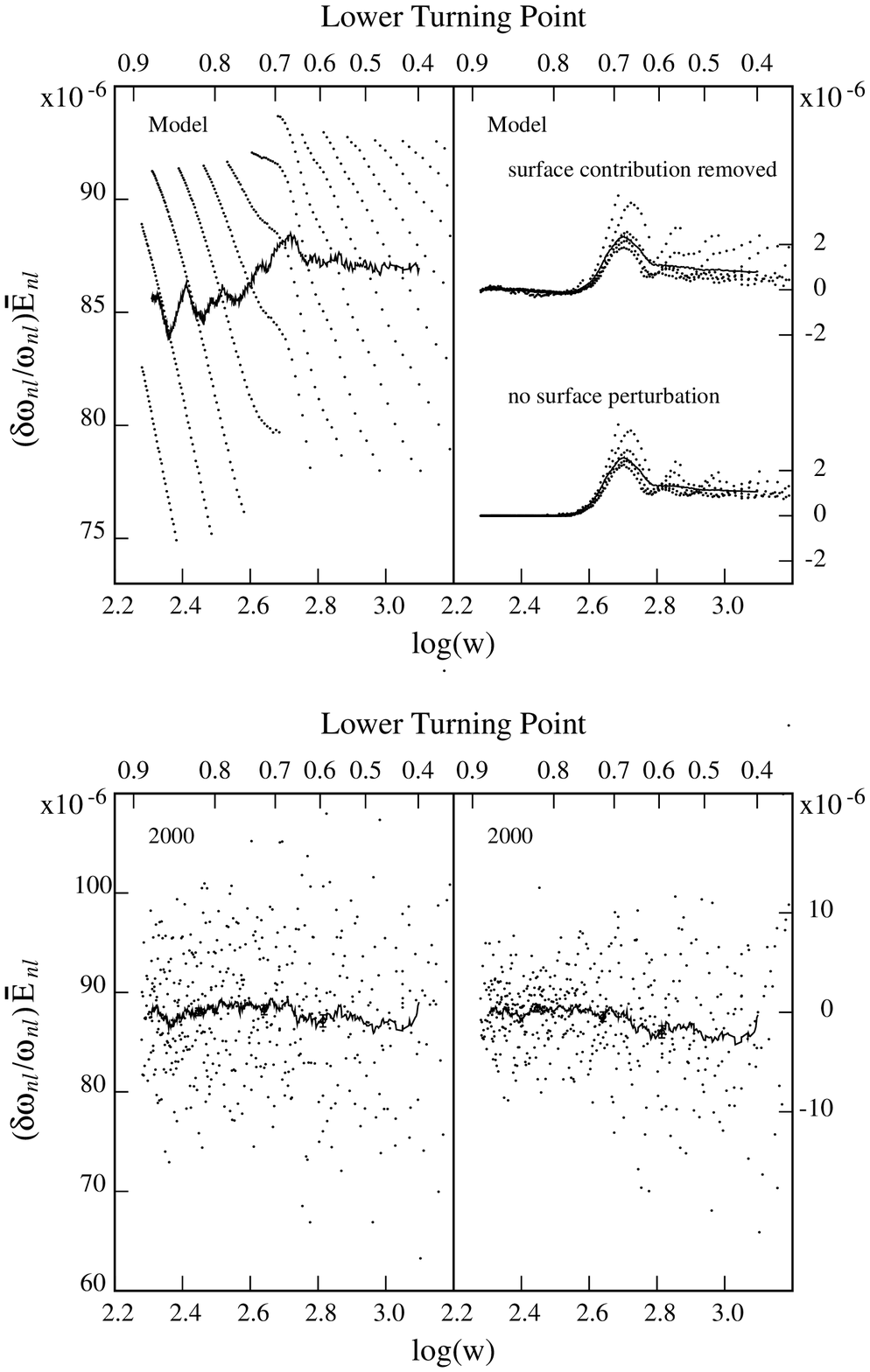}
\figcaption[f1.eps]{
{\it Top}: 
$(\delta \omega_{nl}/\omega_{nl}) \bar{E}_{nl}$ (left panel) and
$[(\delta \omega_{nl}/\omega_{nl}) \bar{E}_{nl}]_{\rm{bcz}}$ (right panel),
computed with models, versus $log(w)$.
The solid lines are the smoothed data using a 41-point box running mean.
For comparison, 
$(\delta \omega_{nl}/\omega_{nl}) \bar{E}_{nl}$ computed with 
only the BCZ perturbation is shown in the lower part of the right panel.
{\it Bottom}:
$(\delta \omega_{nl}/\omega_{nl}) \bar{E}_{nl}$ (left panel) and
$[(\delta \omega_{nl}/\omega_{nl}) \bar{E}_{nl}]_{\rm{bcz}}$ (right panel),
difference between year 2000 and solar minimum, versus $log(w)$.
The error bars at $log(w)=2.45$, 2.64, and 2.81 are shown.
\label{fig1}}
\end{figure}

\begin{figure}[t]
\epsscale{0.6}
\plotone{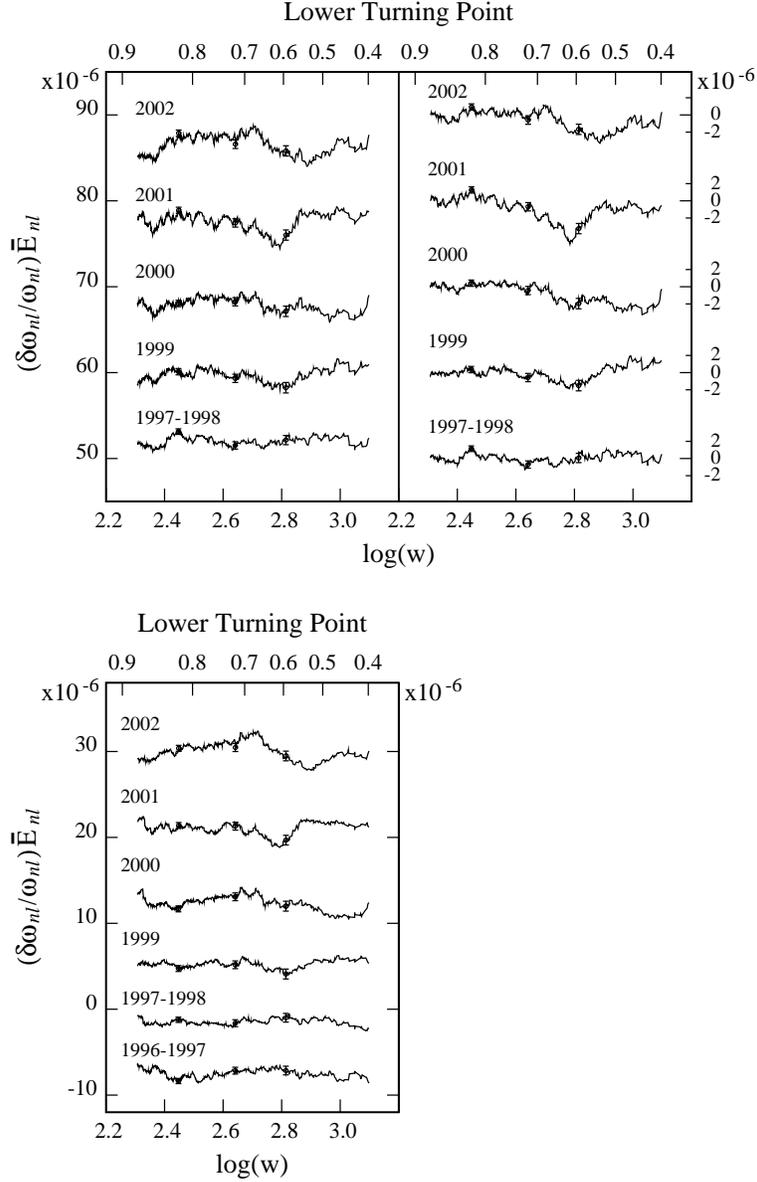}
\figcaption[f2.eps]{
{\it Top}:
Smoothed $(\delta \omega_{nl}/\omega_{nl}) \bar{E}_{nl}$ (left panel) and
$[(\delta \omega_{nl}/\omega_{nl}) \bar{E}_{nl}]_{\rm{bcz}}$ (right panel)
versus $log(w)$ for various periods along the solar cycle,
using the frequency averaged over solar minimum as the reference. 
{\it Bottom}:
Smoothed $(\delta \omega_{nl}/\omega_{nl}) \bar{E}_{nl}$
versus $log(w)$ for various periods along the solar cycle,
using the frequency averaged over all periods as the reference. 
The period of 1996-1997 is 1996.05 - 1997.05. 
The period of 1997-1998 is 1997.07 - 1998.10. 
The curves of smoothed $(\delta \omega_{nl}/\omega_{nl}) \bar{E}_{nl}$
are shifted to avoid interference:
1996-1997 is shifted by $4.5\times 10^{-5}$, 
1997-1998 by $2.5\times 10^{-5}$, 1999 by $-1\times 10^{-5}$, 
2000 by $-2\times 10^{-5}$, and 2001 by $-1\times 10^{-5}$ in both  
top and bottom plots.
The error bars at $log(w)=2.45$, 2.64, and 2.81 are shown.
\label{fig2}}
\end{figure}

We have tried several ways of smoothing: the box running mean,
fit to a polynomial, and Savitzky-Golay smoothing filters \cite{press99}. 
Although different smoothing methods yield different results, 
the main feature discussed here remains similar.  
Here we use a 41-point box running mean to smooth 
$(\delta \omega_{nl}/\omega_{nl}) \bar{E}_{nl}$.
The $w$ dependence of the smoothed 
$(\delta \omega_{nl}/\omega_{nl}) \bar{E}_{nl}$
varies with the solar cycle as shown in Figure \ref{fig2}.
The smoothed $(\delta \omega_{nl}/\omega_{nl}) \bar{E}_{nl}$ 
remains rather flat at low $w$ in all periods, but 
it begins to change (decrease) at $log(w) \approx 2.7$.
The difference between small $w$ and large $w$ 
increases with the magnetic activities on the surface.
The error bars shown in Figures \ref{fig1} and \ref{fig2} are computed 
with the errors of measured frequencies, based on the error propagation. 
It is noted that 
the fluctuation of smoothed $(\delta \omega_{nl}/\omega_{nl}) \bar{E}_{nl}$ 
in the range of $log(w)<2.7$ is approximately the same for all periods
and smaller than the temporal variations at $log(w)>2.7$.
Together with the small error bars of 
smoothed $(\delta \omega_{nl}/\omega_{nl}) \bar{E}_{nl}$, 
it suggests that the random errors of measured frequency can not 
account for the temporal variations at $log(w)>2.7$. 
We can not rule out the possibility of systematic errors,
though it is not clear what kind of systematic error would cause 
this phenomenon.
After removing the near-surface contributions, the relative decrease   
at $log(w)>2.7$ becomes more apparent.  

\begin{figure}[t]
\epsscale{0.7}
\plotone{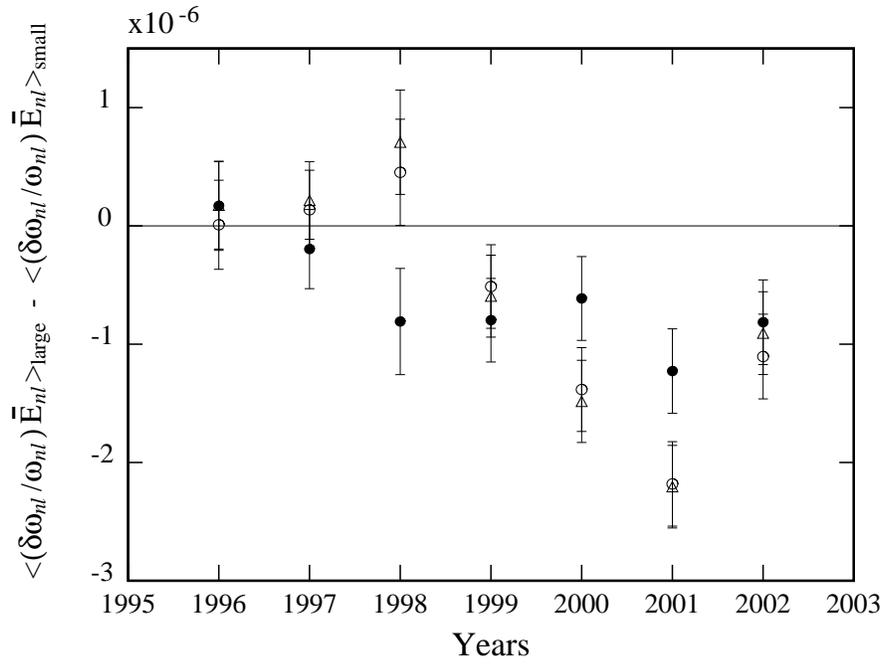}
\figcaption[f3.eps]{
Difference between averages over small $w$ and large $w$ versus time for
$(\delta \omega_{nl}/\omega_{nl}) \bar{E}_{nl}$ (filled circle) and
$[(\delta \omega_{nl}/\omega_{nl}) \bar{E}_{nl}]_{\rm{bcz}}$ (open circle). 
The triangles denote the result for
$[(\delta \omega_{nl}/\omega_{nl}) Q_{nl}]_{\rm{bcz}}$.
It is noted that there are only seven-month data in 1998.
\label{fig3}}
\end{figure}

Negative $[(\delta \omega_{nl}/\omega_{nl}) \bar{E}_{nl}]_{\rm{bcz}}$ 
at $log(w)>2.7$ correspond to the existence of a region near the BCZ
with a smaller wave speed relative to solar minimum.  
From model computations, a decrease in sound-speed
near the BCZ would create a dip in 
$[(\delta \omega_{nl}/\omega_{nl}) \bar{E}_{nl}]_{\rm{bcz}}$ at 
$log(w) \approx 2.7$.  
Because of the small S/N, it is difficult to discuss the existence
of this dip in Figure \ref{fig2}.
However, we can increase the S/N by computing the difference between 
$[(\delta \omega_{nl}/\omega_{nl}) \bar{E}_{nl}]_{\rm{bcz}}$ averaged
over small $w$ and large $w$.  
Figure \ref{fig3} shows the difference between
$[(\delta \omega_{nl}/\omega_{nl}) \bar{E}_{nl}]_{\rm{bcz}}$ averaged
over $log(w)=2.4 - 2.65$ and $log(w)=2.7-3.0$ versus time.
This difference approximately increases with solar activity. 
The cause of the relative decrease in frequency at large $w$ is not clear.  
If magnetic fields at the BCZ causes an increase in wave speed,
the phenomenon observed here might suggest that 
variations of magnetic flux near the BCZ 
and surface magnetic flux are not in phase.

In this study, we use the frequency at solar minimum as the reference and 
investigate solar-cycle variations of frequency change relative to this 
reference frequency.  This approach suffers from the risk that the 
error in frequency at minimum affects the results in all other periods.   
To avoid it, we also use the frequency averaged over all periods 
as the reference.  
The smoothed $(\delta \omega_{nl}/\omega_{nl}) \bar{E}_{nl}$, 
are shown in the lower left panel of Figure \ref{fig2}.  
Although these curves are smoother than those using the frequency
averaged over minimum as the reference,
the relative temporal variations of two groups of curves are consistent. 

Since $S_{nl}$ is proportional to the mode mass $E_{nl}$,
in this study we replace $S_{nl}$ by $E_{nl}/E_{21,0}$.
Although sometimes $Q_{nl}$, which equal $E_{nl}$ normalized by
mode mass of a radial mode of the same frequency,
are used in other studies, the frequency dependence of the mode 
mass of radial modes would make the first term in
equation (\ref{eq:h1234-a}) no longer a function of $w$ only. 
However, in practice we find that using $E_{nl}$ and $Q_{nl}$
yield very close results. 
The difference between 
$[(\delta \omega_{nl}/\omega_{nl}) Q_{nl}]_{\rm{bcz}}$ averaged
over small $w$ and large $w$ using $Q_{nl}$ is also shown in Figure 3.

In removing the near-surface contribution, the choice 
of the range of $w$ in the fit would affect the detailed features of 
$[(\delta \omega_{nl}/\omega_{nl}) \bar{E}_{nl}]_{\rm{bcz}}$.
However, the main feature of smoothed
$[(\delta \omega_{nl}/\omega_{nl}) \bar{E}_{nl}]_{\rm{bcz}}$
is robust with respect to the choice of the fit range.
In deriving equation (\ref{eq:h1-c}) from equation (\ref{eq:h1-a}), 
we have assumed that the perturbation in the CZ is negligible except
near the surface.  If it is not the case, we need to include  
higher-order terms of $w^{-2}$ in equation (\ref{eq:h1-c}).

In summary, Figures \ref{fig1}-\ref{fig2} indicate that
solar-cycle variations of the $w$ dependence of
$(\delta \omega_{nl}/\omega_{nl}) \bar{E}_{nl}$  
at $log(w)<2.7$ and $log(w)>2.7$ are different.  
This phenomenon suggests that the physical conditions in a region near the BCZ 
change with the solar cycle.  
If we attribute the difference in 
$[(\delta \omega_{nl}/\omega_{nl}) \bar{E}_{nl}]_{\rm{bcz}}$ 
between $log(w)<2.7$ and $log(w)>2.7$ to a change in $\Gamma_1$ near the BCZ, 
it corresponds to 
$-\delta \Gamma_1/\Gamma_1(=-\delta c^2/c^2) \approx 2-6 \times 10^{-5}$ 
at $r \approx 0.65 - 0.67 R_{\odot}$, if the FWHM of the Gaussian
perturbation is $0.05 R_{\odot}$.

\acknowledgments

We thank H. Shibahashi, T. Seikii, and M. Takata for helpful discussions 
and suggestions.
We are grateful to S. Basu and A. Kosovichev for comments and criticisms.
We thank Jesper Schou for providing the frequency tables of MDI/SOHO.
SOHO is a project of international cooperation between ESA and NASA.
In \S2, we use J. Christensen-Dalsgaard's solar model and computer codes 
available on his website. 
Authors were supported by NSC of ROC under grant NSC-91-2112-M-007-034.


\clearpage
%
%
%

\begin{thebibliography}{}
%
\bibitem[Antia et al.\ 2001]{ant01}
Antia, H. M. et al. 2001, \mnras, 327, 1029
%
\bibitem[Basu \& Schou\ 2000]{bas00}
 Basu, S. \&  Schou, J. 2000, \solphys, 192, 481
%
\bibitem[Basu \& Antia\ 2002]{bas02}
 Basu, S. \&  Antia, H. M. 2002,
 in Proc. SOHO 12/GONG 2002 Workshop: Local and Global Helioseismology,
 ed. H. Sawaya-Lacoste (ESA SP-517; Noordwijk: ESA), 231 
%
\bibitem[Basu et al.\ 2003]{bas03}
Basu, S., Christensen-Dalsgaard, J., Howe, R., Schou, J., Thompson, M. J.,
Hill, F., \& Komm, R. 2003, \apj, 591, 432
%
\bibitem[Brodsky \& Vorontsov\ 1993]{bro93}
Brodsky, M. A. \& Vorontsov, S. V. 1993, \apj, 409, 455
%
\bibitem[Christensen-Dalsgaard\ 2003]{chr03}
Christensen-Dalsgaard, J. 2003, 
solar model and computer codes, http://bigcat.phys.au.dk/$\sim$jcd/
%
\bibitem[Christensen-Dalsgaard et al.\ 1988]{chr88}
Christensen-Dalsgaard, J., Gough, D. O., \& Perez Hernandez, F. 1988, 
\mnras, 235, 875
%
%
\bibitem[Christensen-Dalsgaard et al.\ 1996]{chr96}
Christensen-Dalsgaard, J. et al. 1996, Science, 272, 1286 
%
\bibitem[Di Mauro et al.\ 2002]{mauro02}
Di Mauro, M. P., Christensen-Dalsgaard, J., Rabello-Soares, M. C.,
\& Basu, S. 2002, \aap, 384, 666
%
\bibitem[Eff-Darwich et al.\ 2002]{eff02}
 Eff-Darwich, A., Korzennik, S. G., Jimenez-Reyes, S. J., \& 
Perez Hernandez, F. 2002, \apj, 580, 574
%
\bibitem[Gough \& Vorontsov\ 1995]{gou95}
Gough, D. O. \& Vorontsov, S. V. 1995, \mnras, 273, 573 
%
\bibitem[Howe et al.\ 1999]{how99}
 Howe, R., Komm, R., \& Hill, F. 1999, \apj, 524, 1084 
%
\bibitem[Libbrecht \& Woodard \ 1990]{lib90}
 Libbrecht, K. G. \& Woodard, M. F. 1990, \nat, 345, 779
%
\bibitem[Press et al.\ 1999]{press99}
Press, W. H., Teukolsky, S. A., Vetterling, W. T., \& 
Flannery, B. P. 1999, Numerical Recipes in C, 2nd ed. 
(Tucson: Cambridge Univ. Press), p. 650
%
\bibitem[Scherrer et al.\ 1995]{sch95}
 Scherrer, P. H. et al. 1995, \solphys, 160, 237
%
\bibitem[Shibahashi\ 1991]{shi91}
 Shibahashi, H. 1991, in Lecture Note in Physics, 388,
 eds. D. Gough \& J. Toomre (Berlin: Spinger-Verlag), 101 
\end{thebibliography}
\end{document}